\documentclass[prb,twocolumn,showpacs,floatfix,amsmath,amssymb,superscriptaddress,longbibliography]{revtex4-1}
\usepackage{amsfonts}
\usepackage{stmaryrd}
\usepackage{bbm}
\usepackage{mathrsfs}
\usepackage{tipa}
\usepackage{amssymb}
\usepackage{txfonts}
\usepackage{graphicx}
\usepackage{dcolumn}
\usepackage{epstopdf}
\usepackage[colorlinks,linkcolor=blue,urlcolor=blue,citecolor=blue]{hyperref}
\usepackage{multirow}
\usepackage{subfigure}
\usepackage{url}
\usepackage{upgreek}
\usepackage{booktabs}
\usepackage{appendix}

\begin{document}
\newcommand*{\cm}{cm$^{-1}$\,}

\title{Direct measurement of photoinduced transient conducting state in multilayer 2$H$-MoTe$_2$}%force line break\\

\author{X. Y. Zhou}
\affiliation{International Center for Quantum Materials, School of Physics, Peking University, Beijing 100871, China}

\author{H. Wang}
\affiliation{International Center for Quantum Materials, School of Physics, Peking University, Beijing 100871, China}

\author{Q. M. Liu}
\affiliation{International Center for Quantum Materials, School of Physics, Peking University, Beijing 100871, China}

\author{S. J. Zhang}
\affiliation{International Center for Quantum Materials, School of Physics, Peking University, Beijing 100871, China}

\author{S. X. Xu}
\affiliation{International Center for Quantum Materials, School of Physics, Peking University, Beijing 100871, China}

\author{Q. Wu}
\affiliation{International Center for Quantum Materials, School of Physics, Peking University, Beijing 100871, China}

\author{R. S. Li}
\affiliation{International Center for Quantum Materials, School of Physics, Peking University, Beijing 100871, China}

\author{L. Yue}
\affiliation{International Center for Quantum Materials, School of Physics, Peking University, Beijing 100871, China}

\author{T. C. Hu}
\affiliation{International Center for Quantum Materials, School of Physics, Peking University, Beijing 100871, China}

\author{J. Y. Yuan}
\affiliation{International Center for Quantum Materials, School of Physics, Peking University, Beijing 100871, China}

\author{S. S. Han}
\affiliation{Beijing Academy of Quantum Information Sciences, Beijing 100913, China}

\author{T. Dong}
\affiliation{International Center for Quantum Materials, School of Physics, Peking University, Beijing 100871, China}

\author{D. Wu}
\affiliation{Beijing Academy of Quantum Information Sciences, Beijing 100913, China}

\author{N. L. Wang}
\email{nlwang@pku.edu.cn}
\affiliation{International Center for Quantum Materials, School of Physics, Peking University, Beijing 100871, China}
\affiliation{Beijing Academy of Quantum Information Sciences, Beijing 100913, China}

\begin{abstract}

Ultrafast light-matter interaction has emerged as a powerful tool to control and probe the macroscopic properties of functional materials, especially two-dimensional transition metal dichalcogenides which can form different structural phases with distinct physical properties. However, it is often difficult to accurately determine the transient optical constants. In this work, we developed a near-infrared pump - terahertz to midinfrared (12-22 THz) probe system in transmission geometry to measure the transient optical conductivity in 2$H$-MoTe$_2$ layered material. By performing separate measurements on bulk and thin-film samples, we are able to overcome issues related to nonuniform substrate thickness and penetration depth mismatch and to extract the transient optical constants reliably. Our results show that photoexcitation at 690 nm induces a transient insulator-metal transition, while photoexcitation at 2 $\mu$m has a much smaller effect due to the photon energy being smaller than the band gap of the material. Combining this with a single-color pump-probe measurement, we show that the transient response evolves towards 1$T^{'}$ phase at higher flunece. Our work provides a comprehensive understanding of the photoinduced phase transition in the 2$H$-MoTe$_2$ system.

\end{abstract}


\maketitle

\section{Introduction}

Layered transition-metal dichalcogenides (TMDCs) are a class of two-dimensional (2D) materials that exhibit a diverse range of physical phenomena \cite{Mak2013,Palummo2015,Sun2016,Fei2018}. MoTe$_2$ is of particular interest due to its ability to crystallize into various structures, including the trigonal prismatic coordinated hexagonal 2$H$ phase (shown in Fig.\ref{Fig:1} (a)-(c)), distorted octahedral coordinated monoclinic 1$T^{'}$ phase, and orthorhombic $T_d$ phase.These different phases possess significantly distinct properties, with 2$H$-MoTe$_2$ being insulating with a band gap of approximately 1 eV, while 1$T^{'}$ and $T_d$ phases exhibit semimetallic nature \cite{Dawson,Clarke,Nature2017}. A temperature-dependent resistivity of 2$H$-MoTe$_2$ thin film is shown in Fig.\ref{Fig:1} (d). The 1$T^{'}$-MoTe$_2$ phase is centrosymmetric and can be transformed into $T_d$-MoTe$_2$ phase with broken inversion symmetry when cooling below $\sim$250 K \cite{Dawson,Clarke}. $T_d$-MoTe$_2$ phase hosts type-II topological Weyl fermions and is also superconducting below 0.1 K \cite{Sun,Chang2016,Qi2016}. Another interesting aspect is that those 2D materials can be easily exfoliated into flakes or prepared to form mono-or multilayered sheets. Intriguingly, bulk MoTe$_2$ materials undergo significant changes in electronic and photoelectric properties when transformed into layered materials \cite{Song2018,Zheng2020}. The formation of various phases in 2D MoTe$_2$ presents a promising prospect for the deliberate manipulation or regulation of phase transitions. The utilization of ultrashort lasers has demonstrated their efficacy in the management of diverse characteristics in layered materials \cite{Wang2012,Lui2014,Kim2019}. For instance, ultrafast laser pulses can induce a nonthermal phase transition between type-II Weyl semimetal $T_d$ phase and normal semimetal 1$T^{'}$ phase in a sub-picosecond timescale \cite{Zhang2019}.

Insulating 2$H$ phase is the most stable state in MoTe$_2$ and its structure is more different from $T_d$ and 1$T^{'}$ phases. Then, it is expected to be difficult to induce a transition from 2$H$ to 1$T^{'}$ or $T_d$ phases. Nonetheless, recent studies suggest that the phase transition from 2$H$ to 1$T^{'}$ phase could be induced by different techniques, including ultrashort laser or terahertz (THz) pulses in either bulk or single layer MoTe$_2$ samples \cite{Song2016,Nature2017,Li2016,Cho2015,Tan2019a,Peng2020}. Such phase transition is expected to be associated with insulator-metal transition (IMT). However, up to now, there has been a lack of direct conductivity measurement identifying the IMT. Here we present time-resolved terahertz to midinfrared measurements to investigate the effects of photoexcitation at 690 nm (1.8 eV) and 2 $\mu$m (0.62 eV) on 2$H$-MoTe$_2$. By conducting two separate experiments on bulk and thin-film samples, we are able to avoid the problems of nonuniform substrate thickness and penetration depth mismatch, and to accurately determine the optical constants after excitation. Our results show that photoexcitation at 690 nm induces a transient IMT, while photoexcitation at 2 $\mu$m has a minimal effect due to the photon energy being smaller than the band gap of the material. In order to establish possible link between the transient IMT and a stable 1$T^{'}$ state, we performed fluence-dependent single-color pump-probe measurement and found that the transition evolves towards the 1$T^{'}$ phase at higher fluences.

\section {Experiments and Results}

\begin{figure*}[t]
 \centering
 \centering\includegraphics[width=16cm]{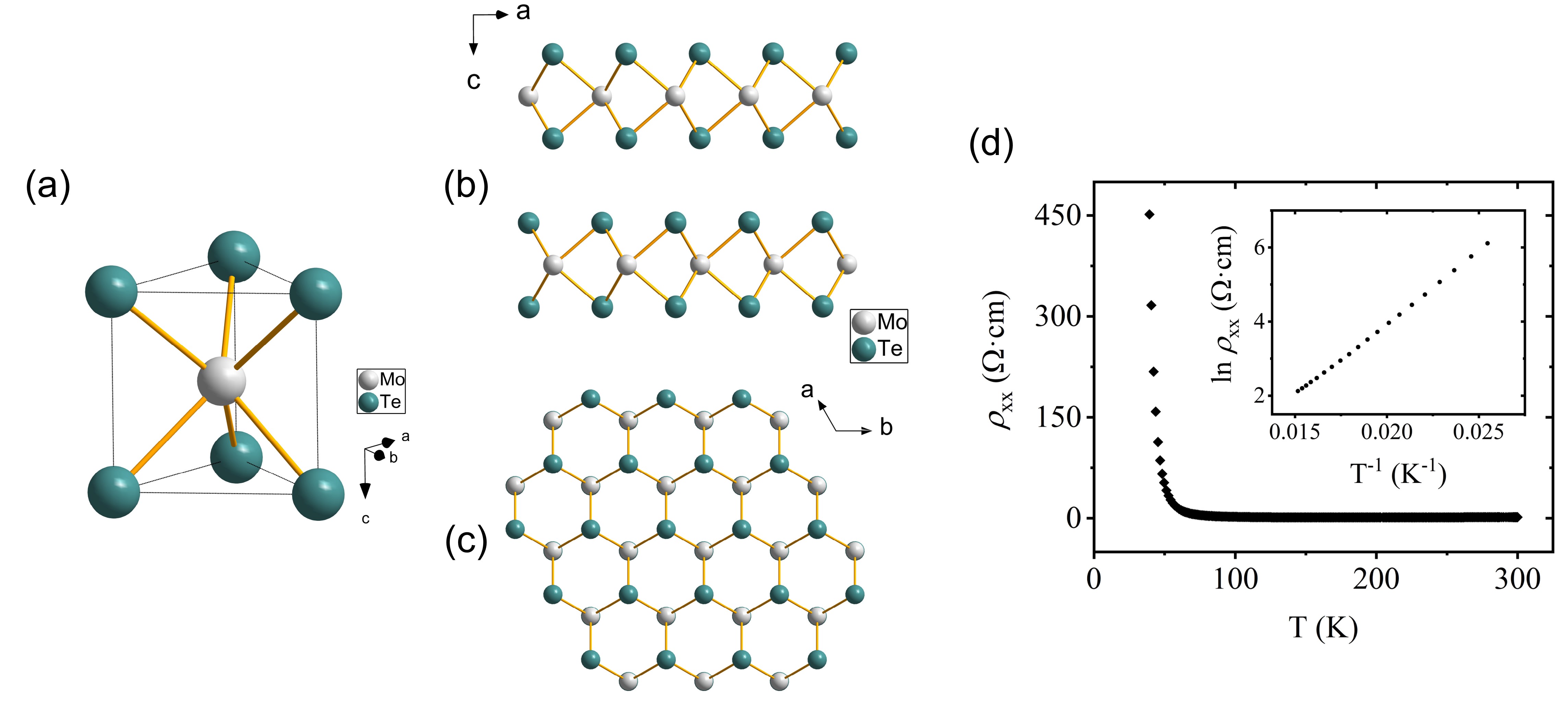}\\
 \caption{\textbf{Lattice structure and the electrical property of 2$H$-MoTe$_2$ phase.} (a) The adjacent Te atoms are connected by thin black lines, forming a triangular prism centered on a Mo atom. (b) and (c) show the inter-plane and in-plane views of the crystal structure, respectively. (d) Temperature-dependent resistivity of a MoTe$_2$ thin film.} \label{Fig:1}
\end{figure*}

We constructed an ultrafast near-infrared pump - terahertz to midinfrared probe system in the frequency range of 12-22 THz. The system uses a Pharos laser from Light Conversion to produce a 1030 nm laser with an energy of 400 $\mu$J at a repetition rate of 50 kHz. 100 $\mu$J of this laser is directed to an optical parametric amplifier (OPA) to generate 800 nm light with a pulse width of 10 fs. The remaining 300 $\mu$J is directed to OPA-twins to produce near-infrared (NIR) or midinfrared (MIR) radiation with tunable wavelength as the pump light. The 800 nm laser pulse is directed to a z-cut GaSe crystal to generate THz radiation in the frequency range of 12-22 THz. The optical path is designed to ensure that the pump pulse and probe pulse reach the sample at the same time, and the probe pulse reaches the probe crystal at the same time as the sampling pulse. A detailed setup of the optical system is provided in the Fig. S1 of appendix.

Transmission time-domain spectroscopy is a well-established method for obtaining equilibrium and photoexcited optical constants. This method directly measures the transient conductivity in the frequency range being studied, providing a clear insight into the dynamic evolution with time delay after photoexcitation \cite{Nuss1987,Hoffmann2009,Ordu2017,Zhang2019}. For insulating materials, a relatively thick sample could be used in the transmission experiment to determine the optical constants in equilibrium state \cite{Shi2018,PhysRevB.104.144408}. The electric field of a THz or MIR pulse that passes through a sample or the same size aperture as reference is recorded as a function of time delay. The recorded time traces are then Fourier transformed to obtain the frequency dependent complex transmission spectra, which contains both magnitude and phase information. Optical constants can then be determined using the Fresnel formula. However, for optical pump THz or MIR probe measurements, it becomes challenging to precisely determine the photoinduced change of optical constants when using thick samples, since the penetration depth of the NIR pump pulse is usually much shorter than that of THz or MIR pulses. To avoid the issue of penetration depth mismatch, a thin film sample grown on a substrate can be used for transmission measurement, provided that the pump pulse can completely penetrate the thin film sample. By using an identical substrate as a reference, both the equilibrium and photoexcited optical constants can be determined in the transmission experiment. However, in the frequency range of 12-22 THz, the commonly used substrates such as sapphire, MgO, and LaAlO$_3$ are not transparent due to high reflectivity between TO and LO phonons. Other substrates like silicon wafer can have a big pump-induced signal, which complicates the analysis of the signal contribution from the sample.

\begin{figure*}[t]
	\centering
	\includegraphics[width=15cm]{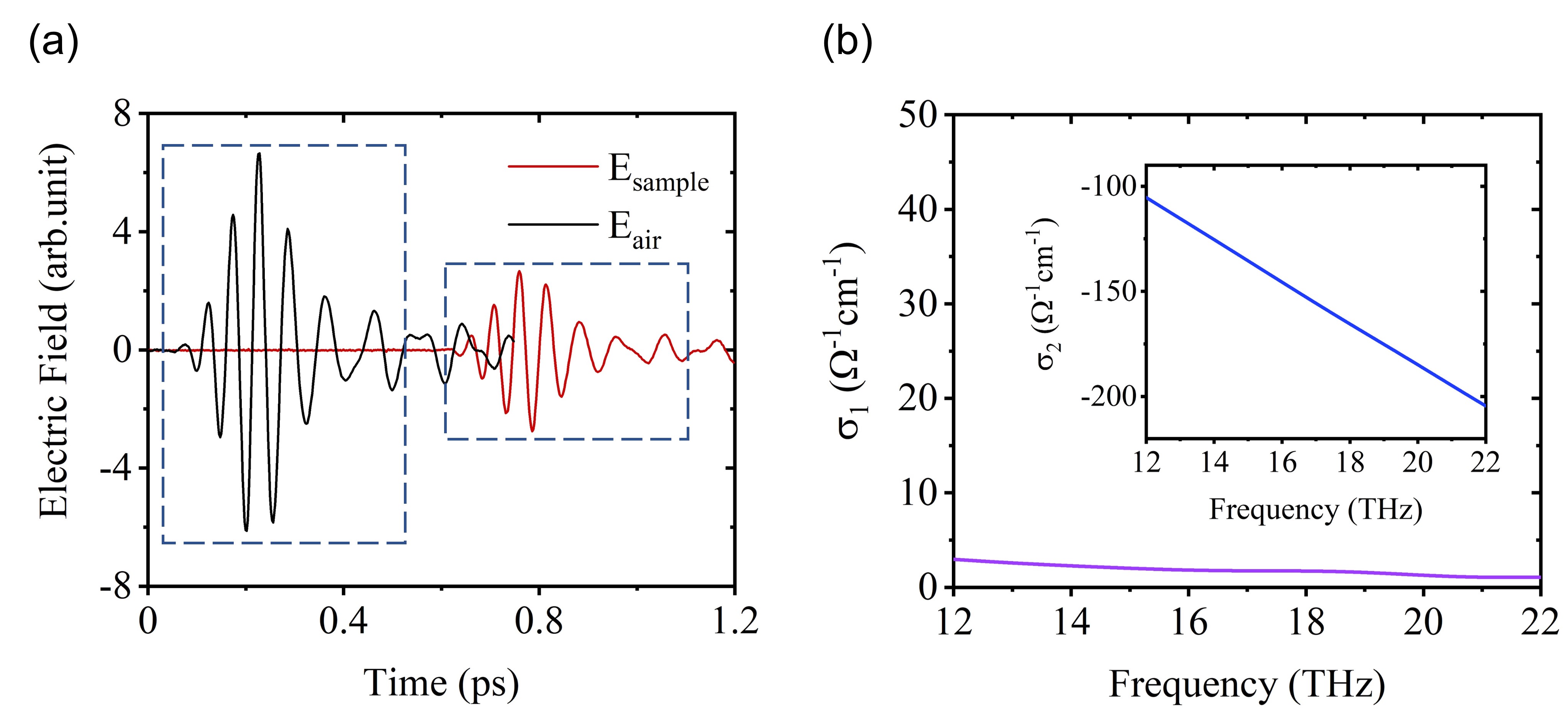}\\
	\caption{\textbf{Static data from a 50 $\mu$m thick flake MoTe$_2$ sample.} (a) The electric field through the air (Black) and through the sample (Red). The shift between the two electric fields in time delay represents the optical path difference causing by the sample. The dashed boxes represent the time window setting during calculation, each window has the same width. (b) Real part of conductivity $\sigma_1$ in the frequency range of 12-22 THz. The inset shows the imaginary part of conductivity $\sigma_2$ in the same frequency range.}\label{Fig:2}
\end{figure*}

During our exploration of various substrates, we discovered that diamond is an ideal substrate for MIR due to its transparency and lack of NIR pump-induced signal. However, the high hardness of diamond made it challenging to obtain uniformly thick polished substrates, resulting in significant heterogeneity that caused errors in determining equilibrium optical constants. Our purchased diamond substrate has a thickness of 0.5 mm, but with a variation of roughly 1 $\mu$m nonuniformity. To overcome these issues, we developed a two-step strategy to detect pump induced optical constants. First, we determined the equilibrium optical constants of 2$H$-MoTe$_2$ using a relatively thick or bulk sample. Second, we work on a thin film of three-layered 2$H$-MoTe$_2$ ($\sim$1.65 nm) being transferred onto the diamond substrate. With the known optical constants in equilibrium, we were able to obtain the pump-induced optical constants without needing to know the thickness of the diamond substrate.

We first present the measurement of optical constants in equilibrium state. The information of the static complex refractive index $\widetilde{n}$ was determined from a flake with 50 $\mu$m thickness, which was obtained by mechanical exfoliation from a MoTe$_2$ single crystal. This flake (bulk) sample was suspended in the air without substrate. The value of $\widetilde{n}$ can be obtained by measuring the probe electric field through the flake sample ($E_{sample}(t)$) and the electric field through the air only ($E_{air}(t)$) \cite{Shi2018,PhysRevB.104.144408}:
\begin{equation}
%\begin{split}
\begin{aligned}
    \label{eq1}
    \centering
    \frac{\widetilde{E}_{sample}(\omega)}{\widetilde{E}_{air}(\omega)} = \frac{4\widetilde{n}exp(\frac{i(\widetilde{n}-1)d\omega}{c})}{(1+\widetilde{n})^2}
%\end{split}
\end{aligned}
\end{equation}
$\widetilde{E}_{sample}$($\omega$) is the the Fourier transform (FFT) of $E_{sample}(t)$, and $\widetilde{E}_{air}$($\omega$) is the FFT of $E_{air}(t)$. $d$ in the formula represents the thickness of the flake sample, $c$ is the speed of light and is a constant, and $\omega$ is the frequency of the probe light. The complex refractive index of air is 1. By knowing the real and imaginary parts of refractive index, one can obtain other optical constants. For example, $\widetilde{\sigma}$ is related to  $\widetilde{n}$ by following equation.
\begin{equation}
%\begin{split}
\begin{aligned}
    \label{eq2}
    \centering
    \widetilde{n}^2(\omega) = 1+i\frac{\widetilde{\sigma}(\omega)}{\varepsilon_0\omega}
%\end{split}
\end{aligned}
\end{equation}
Figure \ref{Fig:2} (a) shows the electric fields passing through the sample in equilibrium state and the air, respectively. The corresponding real and imaginary parts of conductivity of the sample, $\widetilde{\sigma}$=$\sigma_1$+$i\sigma_2$, are shown in Fig.\ref{Fig:2} (b). The very low value of the real part of conductivity in THz to MIR (12-22 THz) reveals that the sample is insulating.

The pump-induced change of optical constants was measured on a three-layer MoTe$_2$ grown by chemical vapor deposition method and transferred onto a diamond substrate. The temperature dependent resistivity is shown in Fig.\ref{Fig:1} (d), which is consistent with the previous report \cite{Lu2022}. The layer thickness is about 1.65 nm, enabling a complete penetration by NIR pulse. $E(t)$ and $E'(t)$ represent the transmitted electric field light before and after pumping, respectively. After pumping, we observe a reduction of peak electric field, indicating a decrease of transmittance or a change to a highly conducting response from the semiconductor ground state. Actually, the pump induced change of electric field ($E'(t)$-$E(t)$=$\Delta$$E(t)$) could be directly measured with much improved signal-to-noise ratio by chopping the pump beam. We verified their equivalence, as shown in Fig.S3 in the appendix. Obviously, the transmitted electric field $E(t)$ is related to the optical constants of the film in equilibrium state and that of the substrate, and $E'(t)$ is related to the optical constants of the film after pumping and also that of the substrate. By taking the Fast Fourier Transform of the transmitted $E'(t)$ and $E(t)$, we can obtain their respective frequency spectra $\widetilde{E}'$($\omega$) and $\widetilde{E}$($\omega$). Note that, for the electric field of the THz pulse passing through the film must consider multiple reflections on both the front and back surfaces of the film. From this, it is easy to derive the ratio of $\widetilde{E}'$($\omega$) and $\widetilde{E}$($\omega$) as,
\begin{equation}
\begin{split}
\label{eq2}
%\begin{aligned}
    \centering
    \frac{\widetilde{E}'(\omega)}{\widetilde{E}(\omega)} &= \frac{exp(i\frac{\widetilde{n}'d\omega}{c})\widetilde{n}'}{exp(i\frac{\widetilde{n}d\omega}{c})\widetilde{n}}\\
                 &\cdot\frac{(1+\widetilde{n})(\widetilde{n}+n_{sub})-(\widetilde{n}-1)(\widetilde{n}-\widetilde{n}_{sub})exp(i\frac{2nd\omega}{c})}{(1+\widetilde{n}')(\widetilde{n}'+\widetilde{n}_{sub})-(\widetilde{n}'-1)(\widetilde{n}'-\widetilde{n}_{sub})exp(i\frac{2\widetilde{n}'d\omega}{c})}
\end{split}
%\end{aligned}
\end{equation}
Here, $\widetilde{n}_{sub}$ is the complex refractive index of the diamond substrate, which has known values. Since the complex refractive index of the material in equilibrium state $\widetilde{n}$ is determined from Eq.(\ref{eq1}), the complex refractive index of the film sample after pumping, $\widetilde{n'}$, can be obtained from Eq.(\ref{eq2}). By knowing the real and imaginary parts of $\widetilde{n'}$, one can obtain any other optical constants.

With the above measurement method we do not need to move the sample position during the measurement. The thickness information of substrate is not involved, therefore, the influence of non-uniformity is avoided. In the meantime, the usage of a fully permeable thin film sample to the pump pulse eliminates the effect of penetration depth mismatch. This measurement method greatly improves the credibility of experimental data. Our results is highly reproducible. It should be noted that, the band structures of 2D materials may change with thickness of sample. For semiconducting 2$H$-MoTe$_2$, only the monolayer has a direct bandgap of 1.10 eV, while the mutilayer and bulk forms have an indirect bandgap of about 1.0 eV \cite{Ruppert2014}. Therefore the electronic structure of the trilayer sample is more similar to the bulk sample. Two recent experimental studies \cite{Fang2022,Jung2022} discussed the thickness dependence of the optical properties. Jung et al. \cite{Jung2022} suggested that there may be as much as a factor of two difference between the 3-layer conductivity and the bulk conductivity at 1.8 eV, while Fang et al. \cite{Fang2022} reported much weaker dependence of $\epsilon_2$ on thickness. We noted that both investigations presented the optical constants only in the high energy scale. Their difference becomes insignificant in the low frequency. As can be seen in Fig. 2 (a) of Jung et al.\cite{Jung2022}, the conductivity tends to converge below 0.8 eV. Since we are focusing on the pump-induced changes in the MIR to THz region, we believe that the effect is minimal.

\begin{figure*}[htbp]
	\centering
\includegraphics[width=17.5cm, trim=0 0 0 0,clip]{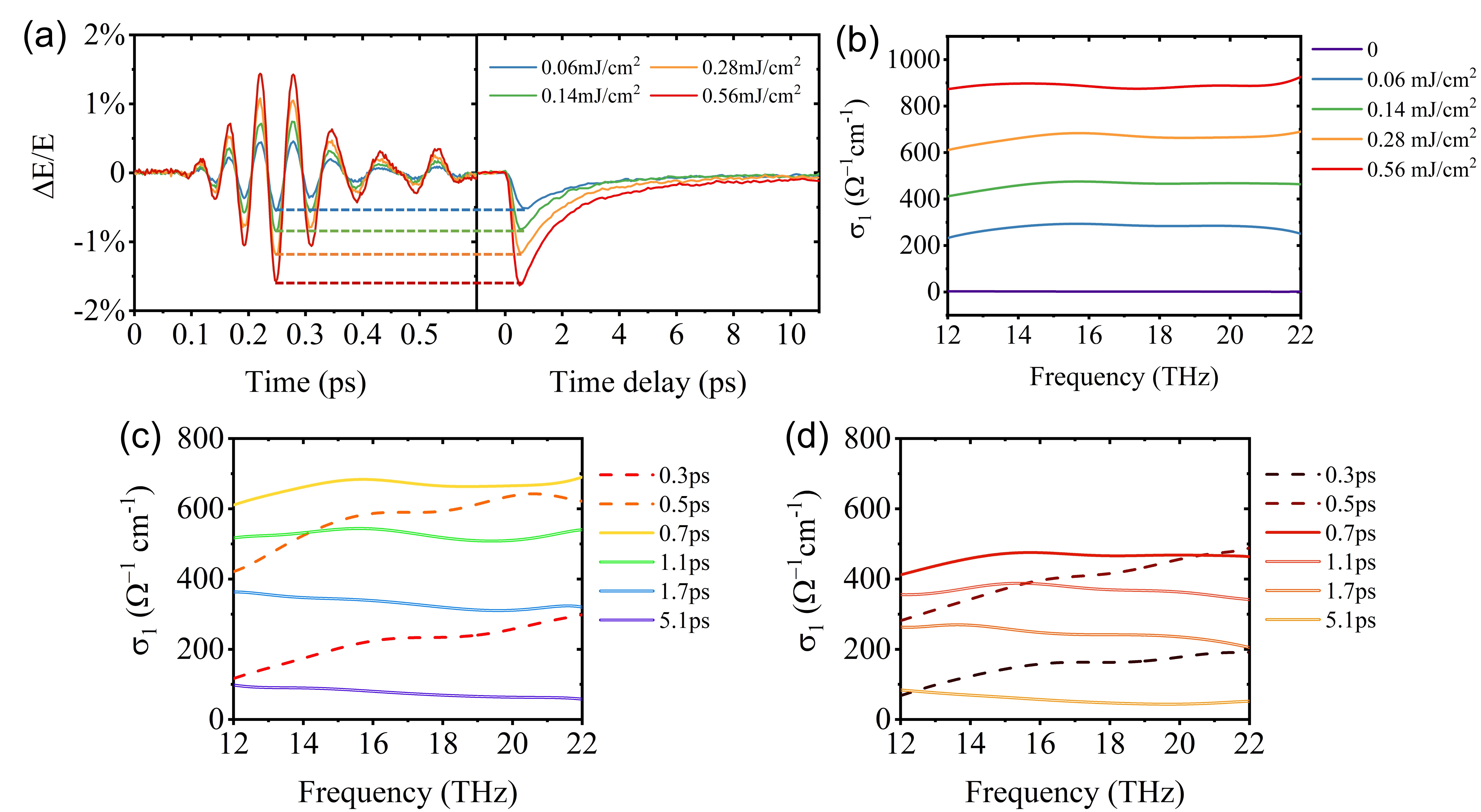}\\

	\caption{\textbf{Near-infrared pump - terahertz to midinfrared probe trasmission spectra.} (a) Right panel: Pump-probe signal with time delay of a three-layer 2$H$-MoTe$_2$ film at several different pump fluences. Left panel: Relative change in terahertz to midinfrared probe electric field. (b) The real part of conductivity at the pump-probe peak position in different fluences. The bottom curve is static conductivity determined from the thin flake sample. (c) and (d) show the time evolution of the real part of conductivity measured at fluences of 0.28 mJ/cm$^2$ and 0.14 mJ/cm$^2$, respectively. Dash curves and solid curves represent the spectra in the rising edge, peak and falling edge of pump-probe time delay, respectively.} \label{Fig:3}
\end{figure*}

Figure \ref{Fig:3} shows our measurement results on a three-layer thin film sample at room temperature. Figure \ref{Fig:3} (a) displays the pump-probe signal excited by NIR pulse of 690 nm at several different fluences (right panel). The signal reaches a sharp apex within 0.7 picosecond (ps) after the pump pulse coming, and decays to a small, non-zero value which maintains at least a few picoseconds. The left panel shows the relative change in MIR probe electric field at the peak position. With only three unit cells (1.65 nm), the relative change reaches 1.7$\%$ at a fluence of 0.56 mJ/cm$^2$. This represents a prominent effect. Indeed, the conductivity spectra derived in this frequency range exhibit significant change, as shown in Fig.\ref{Fig:3} (b). The bottom curve in this plot is from measurement on thin flake sample without external excitation, which has very small conductivity values. We observe that, after pumping, the conductivity increases by more than two orders of magnitude. With increasing fluence, the conductivity increases continuously. The dramatic increase provides direct evidence for a photoinduced transient conducting state. The peak value of $\Delta$$E(t)$/$E(t)$ and extracted conductivity saturate with fluence, but the time dependence looks fairly similar at all fluences shown in the figure. Similar behavior was reported by Sahota et al., on pump-induced reflectance change in visible energies \cite{Sahota2019}.

Figure \ref{Fig:3} (c) and (d) show the conductivity spectra at different time delays after excitation by two different fluences 0.28 mJ/cm$^2$ and 0.14 mJ/cm$^2$. The time-zero is defined at the point when the pump light just reaches the sample, that is, where the pump-probe signal occurs. Then, 0.3 ps and 0.5 ps belong to the rising edge, 0.7 ps corresponds to the peak position of the pump-probe signal, and any time delay after 0.7 ps belongs to the falling edge. During the rise of the signal, the value of $\sigma_1$ increases, reaches its maximum at the top of the signal, then decreases with time delay after the peak. In addition, an intriguing phenomenon is observed. We find that, in the rising edge, $\sigma_1$ increases with increasing frequency, while a slight decreasing behavior is seen in the falling edge. This phenomenon was observed in both pumping fluences. The results may suggest formation of a peak in conductivity beyond the measured frequency in the rising edge, followed by a gradual transformation of the spectral weight from high to low frequencies over time. This is related to the fact that, when the pump pulse first arrives, the sample is mostly in a semiconductor state. We notice that there could be a 0.3 ps time variation across the probe spot due to the non-normal incidence of pump beam, but that cannot explain the systematic evolution feature, since all measurements at different pump-probe time delays should contains the same time variation across the probe spot.

\begin{figure*}[t]
 \centering
 \centering\includegraphics[width=15cm]{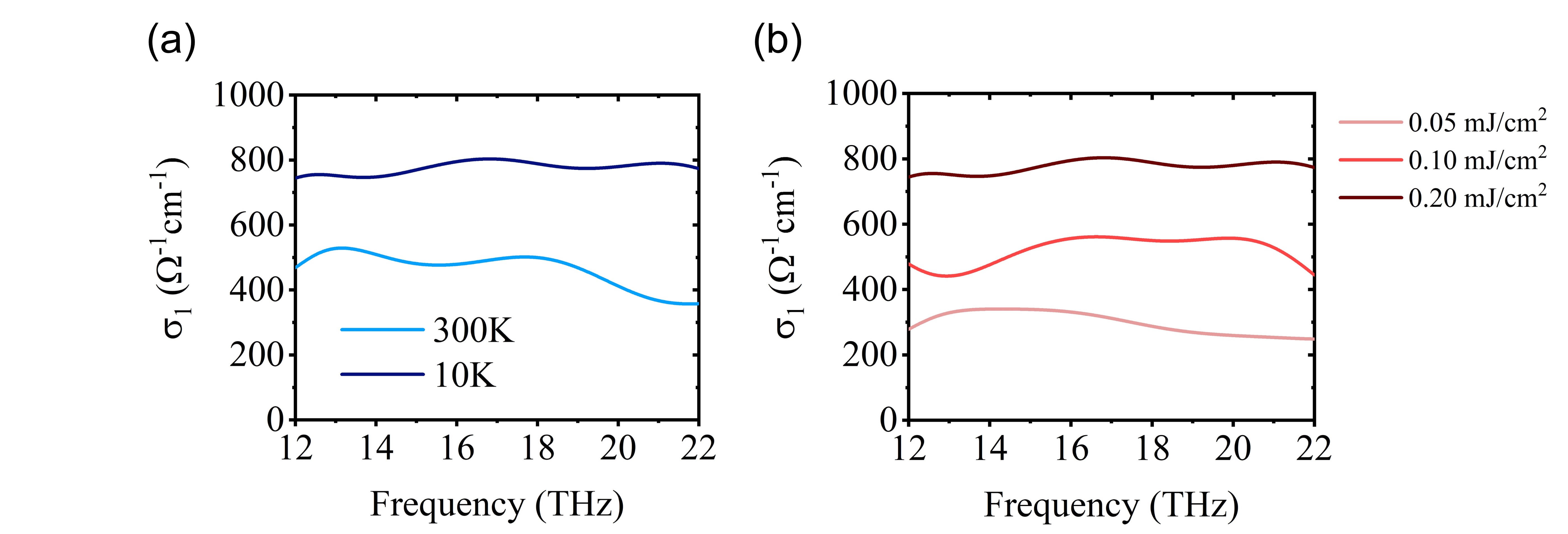}\\
 \caption{\textbf{The photoinduced conductivity at 300 K and 10 K.} (a) The real part of conductivity $\sigma_1$ at 10 K and 300 K at the pump-probe peak position with the 690 nm pump pulse excitation. (b) The real part of conductivity $\sigma_1$ at 10 K under three different pump fluences.} \label{Fig:4}
\end{figure*}

To see how the photoinduced effect changes with temperature, we measured the spectral change at two different temperatures. Figure \ref{Fig:4} (a) shows that $\sigma_1$ at the peak position after pumping with a fluence of 0.20 mJ/cm$^2$ at 300 K and 10 K. Note that, due to the added cryostat window, the actual power is difficult to measure accurately. The results indicate a more significant enhancement in transient conductivity at lower temperatures, associated with reduced photocarrier scattering. Additionally, we performed measurements at 10 K using different pump fluences, as shown in Fig.\ref{Fig:4} (b). The results reveal that the light-induced transient conductivity increases with higher fluence, much like the room temperature case.

The measurements conducted above in this study were carried out using a pump pulse with a wavelength of 690 nm, which has an energy of 1.8 eV. This energy is significantly higher than the energy gap of 2$H$-MoTe$_2$, which is about 1 eV. To investigate the effects of a pump pulse with lower energy than the band gap, measurements were also performed using a pump pulse with a wavelength of 2 $\mu$m (0.62 eV). The results, shown in Fig. S4 in the appendix, reveal that a pump-probe signal can still be resolved even if the pump light energy is lower than the band gap. This may be due to a multiphoton effect. However, the value of the signal under 2 $\mu$m pump pulse is much lower than that under 690 nm pump. Furthermore, the signal from the three-layer sample pumped by 2 $\mu$m is too weak to accurately calculate the optical constant.

We also tried measurement on a relatively thick thin film, about 55 nm, transferred onto the diamond substrate, and showed the results in Fig. S5 in the appendix. We found that the extracted optical conductivity has smaller values than the 1.65 nm (three unit cells) thin film sample. This is because the pump pulse can not completely penetrate the thick film sample, and the detected THz signal can pass the region being not excited by the pump pulse.

\begin{figure*}
 \centering
 \centering\includegraphics[width=16cm]{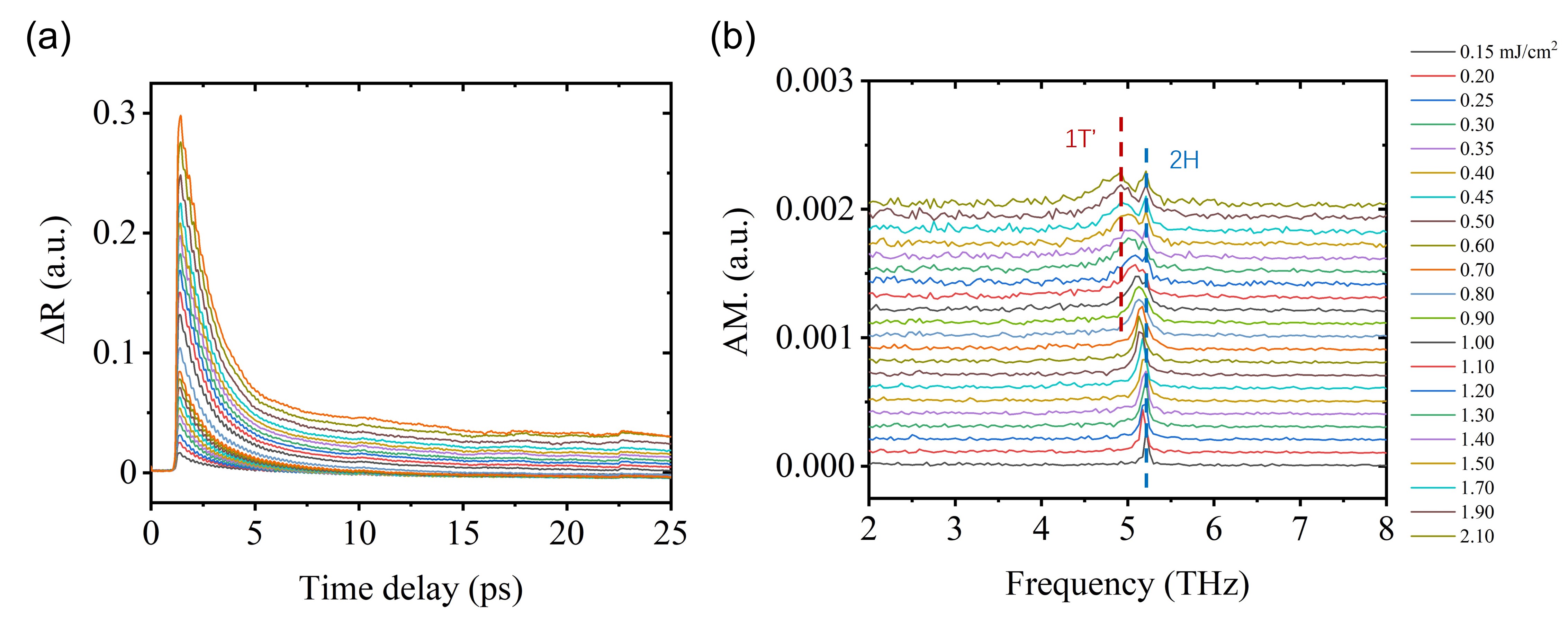}\\
 \caption{(a) 800 nm pump-800 nm probe signal of 2$H$-MoTe$_2$. (b) FFT of the oscillations after subtracting the background signal. Two coherent phonon frequencies are observable: 171.5 cm$^{-1}$ (2$H$ phase), 167.5 cm$^{-1}$ (1$T'$ phase).} \label{Fig:5}
\end{figure*}

Our above experiments represent a direct probe of the transient conductivity in the frequency range of 12-22 THz of a three-layer 2$H$-MoTe$_2$, demonstrating that NIR ultrafast pulse can be effective to trigger highly conducting states in sub-picosecond time scale, whereas photoexcitation at 2 $\mu$m has a tiny/negligible effect due to the photon energy being smaller than the band gap of the material. We address that the transition from insulator to conducting state is transient and returns to the initial state within a few picoseconds. In literature, there are a number of reports showing irreversible transition from 2$H$ to 1$T'$ phase by photoexcitation or even strong field THz pulses \cite{Cho2015,Wang2017a,Krishnamoorthy2018,Peng2020}. We remark that we have used much smaller fluence in the present measurement.

The above time resolved THz to MIR spectral measurement with high pulse repetition rate is not suitable to investigate a nonreversible transition. In order to establish a link between the transient IMT and a stable 1$T'$ state, we also performed single color pump-probe measurement on 2$H$-MoTe$_2$ to look into possible change of coherent phonon modes. We use a 800 nm amplified laser system with a pulse duration of 35 fs and repetition rate of 1 kHz for the experiment. We measured the reflection change on a thin flake sample with a thickness of about 20 nm. Figure \ref{Fig:5} shows reflection change as a function of time delay at various pump fluences. We can see clearly oscillations in the pump-probe signals of Fig.\ref{Fig:5} (a). After subtracting background and performing FFT, we can see coherent phonons in the frequency-domain of Fig.\ref{Fig:5} (b). At low fluences, only the phonon representing 2$H$ phase (about 171.5 cm$^{-1}$) is present \cite{Nature2017}. With increasing fluence, the peak broadens and gradually splits into two, represents 2$H$ and 1$T'$ phase (about 167.5 cm$^{-1}$) \cite{Nature2017}, respectively. With this observation, we expect that, at sufficient fluence, a stable 1$T'$ phase could be induced as such reported in literature. In an earlier report on few layer 2H-MoTe$_2$, a linear relationship between the fluence and the photoconductivity was observed \cite{Zheng2020}. Such behavior was usually observed in semiconductors pumped at small fluences, which could be attributed to the change in photoexcited carrier density. The present work shows clearly a sublinear fluence dependence. As the pump fluence is much higher than that used in ref. \cite{Zheng2020}, the saturation behavior likely results from the depletion of states due to the excitation of electrons from the valence band to the conduction band, leading to a reduced absorption for photon energies across the band gap.

Our research provides an in-depth examination of the photo-induced phase transition from an insulating to a  highly conducting state in 2$H$-MoTe$_2$. We find that the creation of a sufficient number of excited charge carriers is necessary to trigger a transient phase transition. This is reflected by the fact that photo-excited states appear to be sensitive to the energy of ultrashort laser pulses. A reversible IMT occurs when the energy of the NIR laser pulses surpasses the band gap, even at low fluences. However, when the laser pulse energy is lower than the band gap, the effect is minimal. With increasing fluence, we expect a stable 1$T'$ phase could be observed gradually. We can infer that the reported light-induced conducting state is the result of a much higher pulse fluence or electric field applied to the samples. Our findings offer new insight into the transition process from a transient IMT to a nonthermal and nonreversible 1$T'$ phase at sufficiently high fluence.

\section {Summary}

In this study, we develop a NIR pump-THz to MIR (12-22 THz) probe system in transmission geometry to measure the transient optical conductivity in 2$H$-MoTe$_2$ layered material. With an improved two-step measurement method, we are able to avoid issues with nonuniform substrate thickness and penetration depth mismatch and to accurately extract the photoinduced optical constants in the measured MIR range at different time delays. Our results show an enormous transient increase in the real part of conductivity, yielding direct evidence for the photoinduced transient IMT in 2$H$-MoTe$_2$. Additionally, our findings demonstrate that ultrashort laser pulses with photon energy below the band gap produce a significantly smaller effect. Incorporated with a single-color pump-probe measurement, we demonstrating that NIR ultrafast pulse can be effective to trigger highly conducting states in 2$H$-MoTe$_2$ in sub-picosecond time scale. Our work offers new and deep insight into the photoinduced phase transition in the 2$H$-MoTe$_2$ system.

\section {Appendix}

\begin{appendices}
\section{Fabrication of samples}

The 2H-MoTe$_2$ single crystals were synthesized by traditional chemical vapor transportation technology. The quartz tube loaded with the elements (99.99 $\%$ Mo power and 99.999 $\%$ Te pellets) in a stoichiometric ratio together with additional iodine (6 mg/cm$^3$) as the transport agent was vacuumed and flamed-sealed. Then the quartz tube was heated at 750$^0$ C (cold zone) and 850$^0$ C (hot zone) for 2 weeks in a two-zone furnace, followed by a slow cooling process (2$^0$ C/h) to room temperature. Thin flake samples could be easily obtained by mechanical exfoliation from the bulk 2H-MoTe$_2$ single crystal. A flake of 50 $\mu$m thickness was used to get equilibrium data. The sample of 55 nm thickness was obtained by further mechanical stripping from MoTe$_2$ flake. Their thickness was measured by a step meter.

\begin{figure*}
	\centering
	\includegraphics[width=12cm]{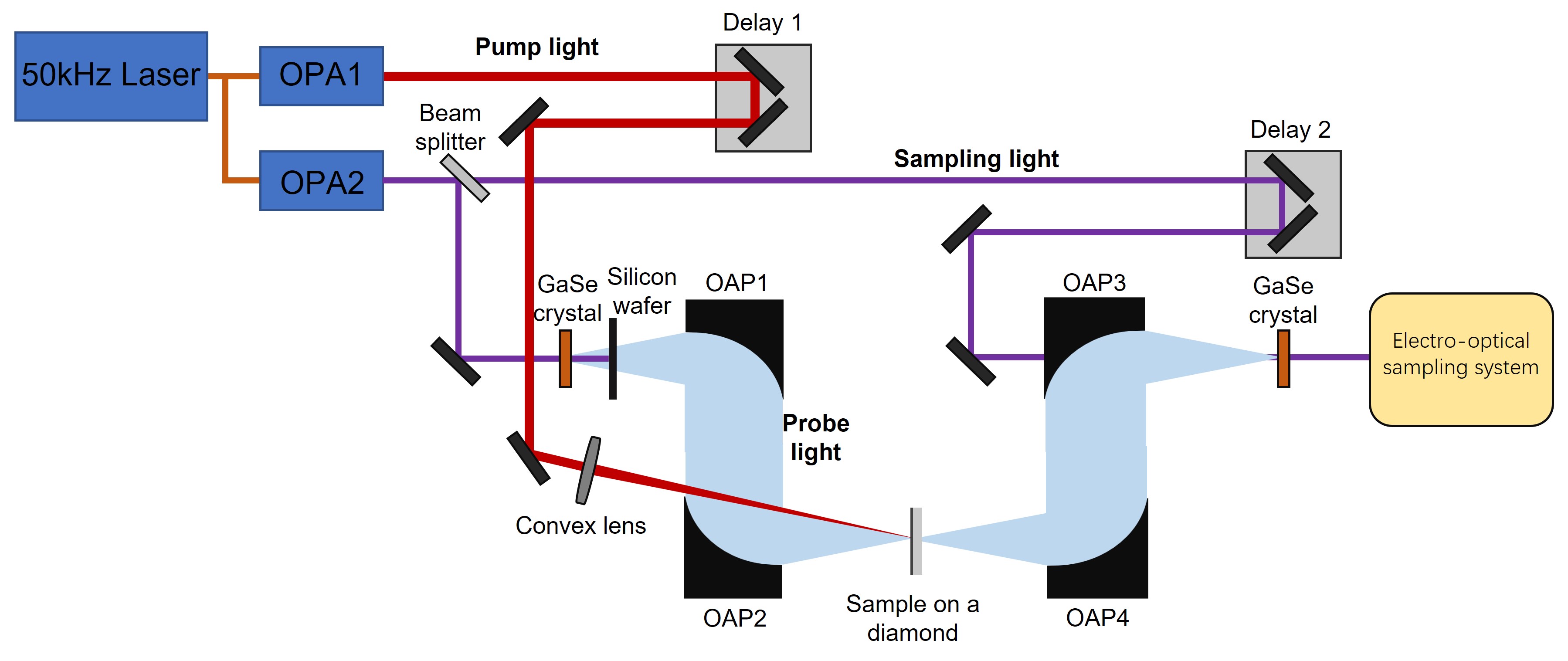}\\
	Fig. S1: The schematic diagram of the optical-pump THz to MIR-probe trasmission spectral system.
\end{figure*}

The three-layer 2H-MoTe$_2$ sample is completely covered on a 5 mm $\times$ 5 mm diamond substrate. We provide diamond substrate to SixCarbon technology company. They transferred the film onto the diamond substrate. The three-layer film of 2H-MoTe$_2$ with the thickness of 1.65 nm is grown on the SiO/Si substrate by CVD. Then, the PMMA solution is added and heated to cure the PMMA into a film, the PMMA film is stick to the three-layer MoTe$_2$ sample film. After the sample film bonded to the diamond substrate by the viscous PMMA film, the PMMA is washed off by acetone solution.

\begin{figure}[h]
	\centering
	\includegraphics[width=7cm]{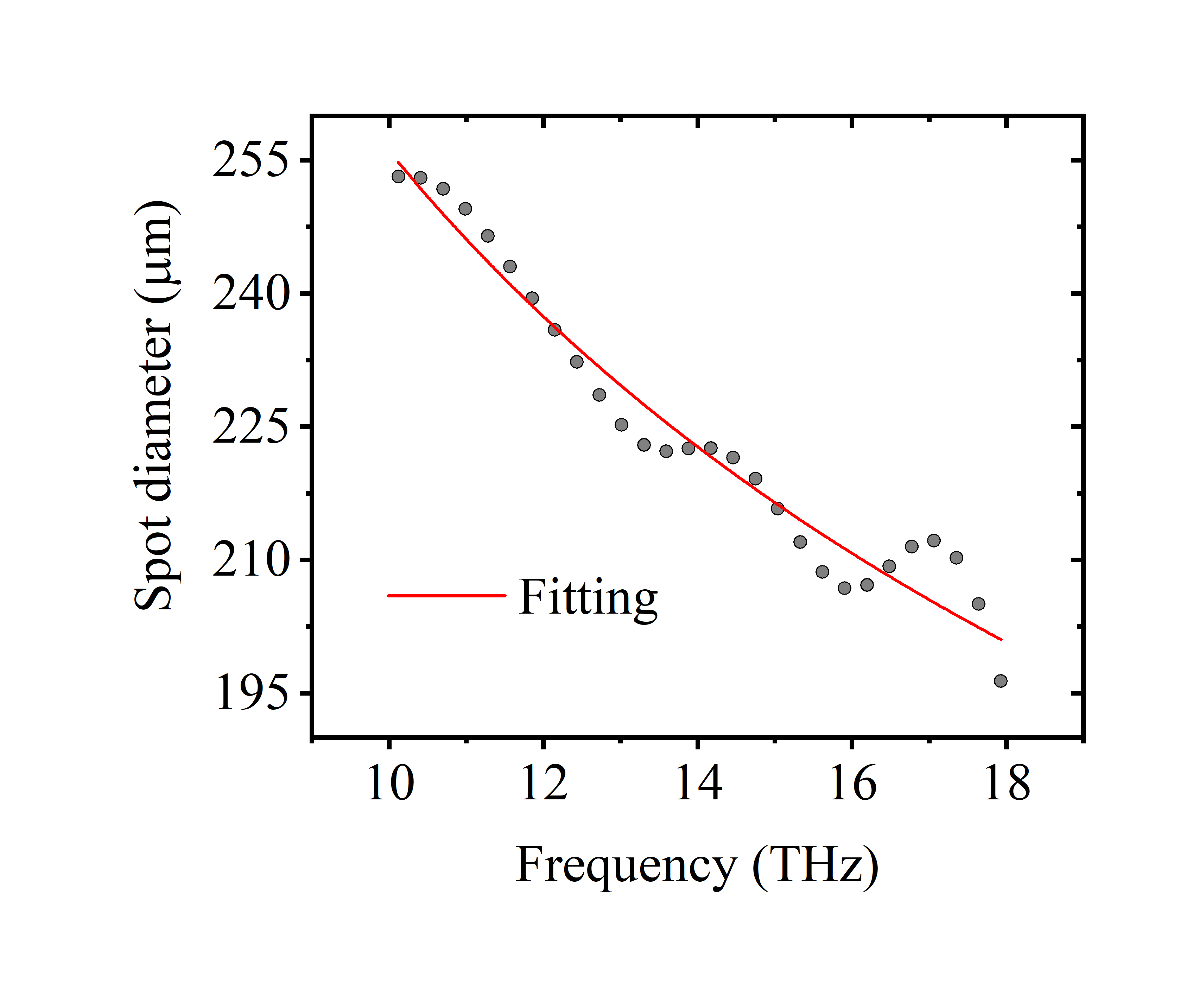}\\
	Fig. S2: The spot diameter of probe light.
\end{figure}

\section{Experimental system}

The schematic diagram of the experimental system is shown in Fig.S1. There are three light beams in the system, namely pump, probe and sampling beams, all of them have a repetition rate of 50 kHz. The pump light pulses (690 nm and 2 $\mu$m) have a diameter of 330 $\mu$m, generated by an optical parametric amplifier (OPA1). The probe light is produced in a GeSe crystal by a 800 nm laser (generated by OPA2), with the spectrum mainly concentrated in 12-22 THz. Both OPAs are pumped by a laser with a central wavelength of 1030 nm and a repetition rate of 50 kHz. The spot size of probe light was measured by knife edge method. As shown in Fig.S2, the diameter of 12 THz is 240 $\mu$m. The pump light is larger in diameter than the probe light to ensure when the two beams coincide, the region measured by the probe beam is fully excited. Probe light is normal incident on the sample, and the transmitted light that passes through the sample and the substrate is incident on the detection GaSe crystal with a sampling light. The sampling beam enters the electro-optical sampling system \cite{Sell08} and picked up by balancing detectors after modulated by the probe light on the detection crystal. There is an angle of 23$^\circ$ between the pump and the probe beams. The relative change of probe is the ratio of the electric field variation ($\Delta$$E(t)$) after pumping to static electric field ($E(t)$). The pump beam and the sampling path each has a translation stage (Delay 1 and Delay 2). The ultrafast time-domain spectral detection is realized by controlling the movement of the two translation stages. Pump-probe signal is the relative change evolution of pump light oscillation time. The electric field decreases after pumping, so the signals are negative (shown in Fig.3 (a) in the main text). The pump-probe signal can be measured by fixing Dealy 2 and moving Delay 1. $\Delta$$E(t)$ can be measured by moving two delays simultaneously. Similar experiment scheme were reported in earlier time-resolved THz spectroscopy measurements, such as in Reference \cite{Kindt7/7/,Zhang2017}.

\begin{figure}[b]
	\centering
	\includegraphics[width=6cm]{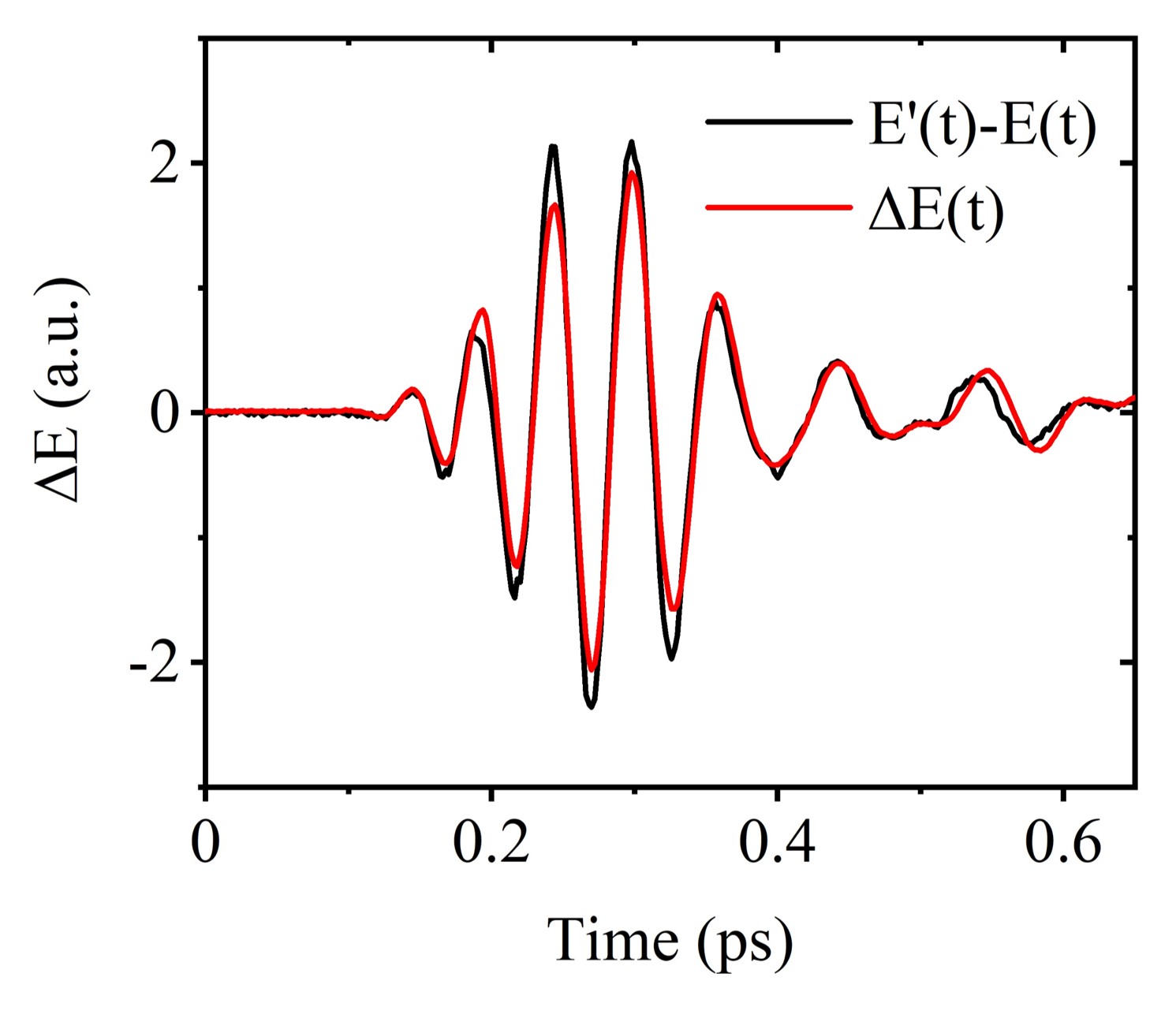}\\
	Fig. S3: A comparison between $\Delta$$E(t)$ and $E^\prime(t)$-$E(t)$.
\end{figure}

Meanwhile, the electric field variation ($\Delta$$E(t)$) is the electric field after pumping $E^\prime(t)$ minus the static electric field $E(t)$. $\Delta$$E(t)$ can be directly measured by chopping the pump beam, or by measuring $E^\prime(t)$ and $E(t)$ separately by chopping the probe beam. In principle, the amplitude and phase of $E^\prime(t)$-$E(t)$ and $\Delta$$E(t)$ should be equal, however, in actual measurement the signal-to-noise ratio will be better by directly measuring $\Delta$$E(t)$. During the experiment, we need to judge the phase of $\Delta$$E(t)$ by $E^\prime(t)$-$E(t)$. The comparison between $\Delta$$E(t)$ and $E^\prime(t)$-$E(t)$ is shown in Fig. S3.

\section{Comparison of varying pump wavelengths}

In addition to the 690 nm pump we used to excite the MoTe$_2$ layers, we also used a 2 $\mu$m pump light as a constrast. We can clearly see in Fig. S4, the signal of 2 $\mu$m pump is far less than 690 nm pump. The signal of 2 $\mu$m shown in Fig. S4 is the result of multiple averages, it is too small to get reliable optical constants. The data was measured at the vertex of the relative change of THz to MIR probe electric field. The effect of pump wavelength is discussed in the text.

\begin{figure}
	\centering
	\includegraphics[width=7cm]{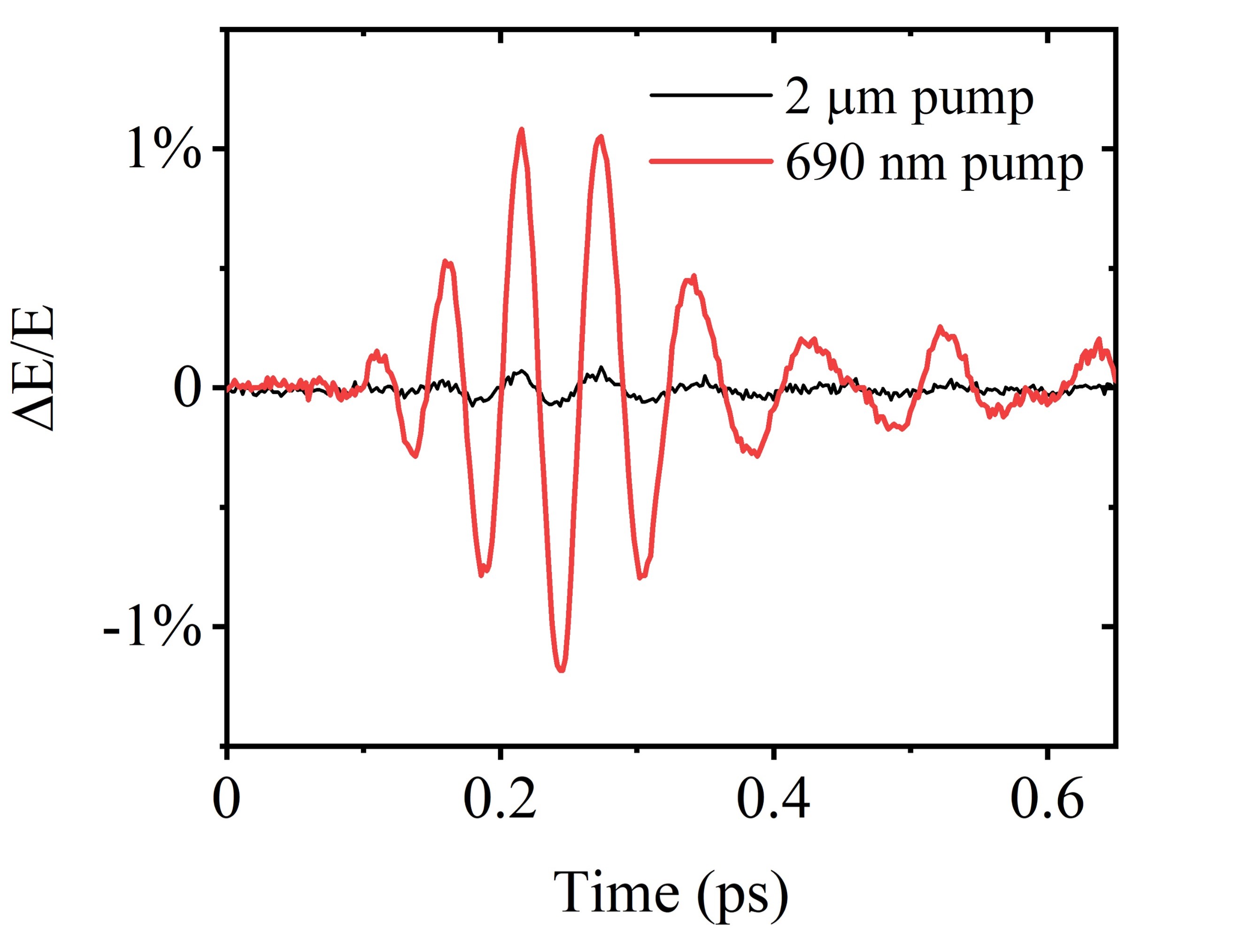}\\
	Fig. S4: The comparison between the signal with 2 $\mu$m pump light (black curve) and the signal with 690 nm pump light (red curve) under the same pump fluence of 0.28 mJ/cm$^2$.

\end{figure}

\section{Effect of pump light penetration depth}

\begin{figure}
	\centering
	\includegraphics[width=8cm]{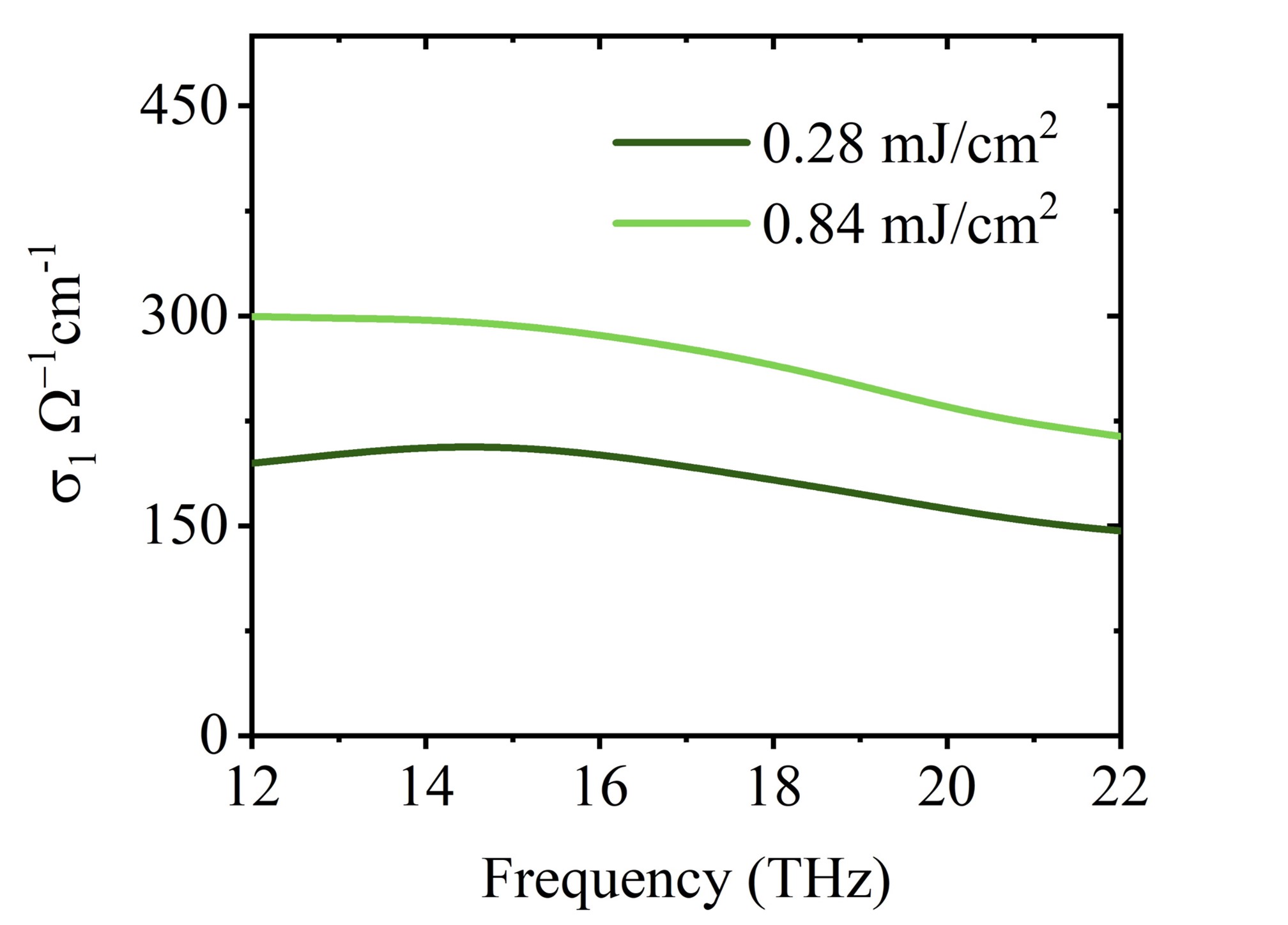}\\
	Fig. S5: The calculated real part of conductivity of the 55 nm thick sample at two different pump fluences.
\end{figure}

We also measured the 690 nm pump-THz to MIR probe signals of a 55 nm thick 2H-MoTe$_2$ sample attached to the diamond. This sample contains around 100 atomic layers and can be considered a bulk material. Even though the pump-probe signal of this sample was much greater than the 1.65 nm film, the change in optical constant calculated by Eq.(3) was much smaller. Compared to the photoconductivity shown in Fig 2 (e) of the main text, the response of the three-layer sample was significantly greater. This is due to the penetration depth, which is influenced by the pump light wavelength and material properties. In this case, 690 nm pump beam could not completely penetrate the bulk sample, but could easily pass through the three-layer 2H-MoTe$_2$. This means that the bulk sample was not fully excited by the pump pulses. The probe beam was able to pass through both samples completely, meaning that the detected region was not completely excited by the pump light if the sample was too thick.

\section{Lifetime of the transient conducting state}

To examine whether the lifetime of the transient conducting state has clear pump fluence dependence, we fitted the pump-probe data in Fig.3(a). The fitting results are shown in Fig. S6. We found that the lifetime does not show significant change.

\begin{figure}
	\centering
	\includegraphics[width=9cm]{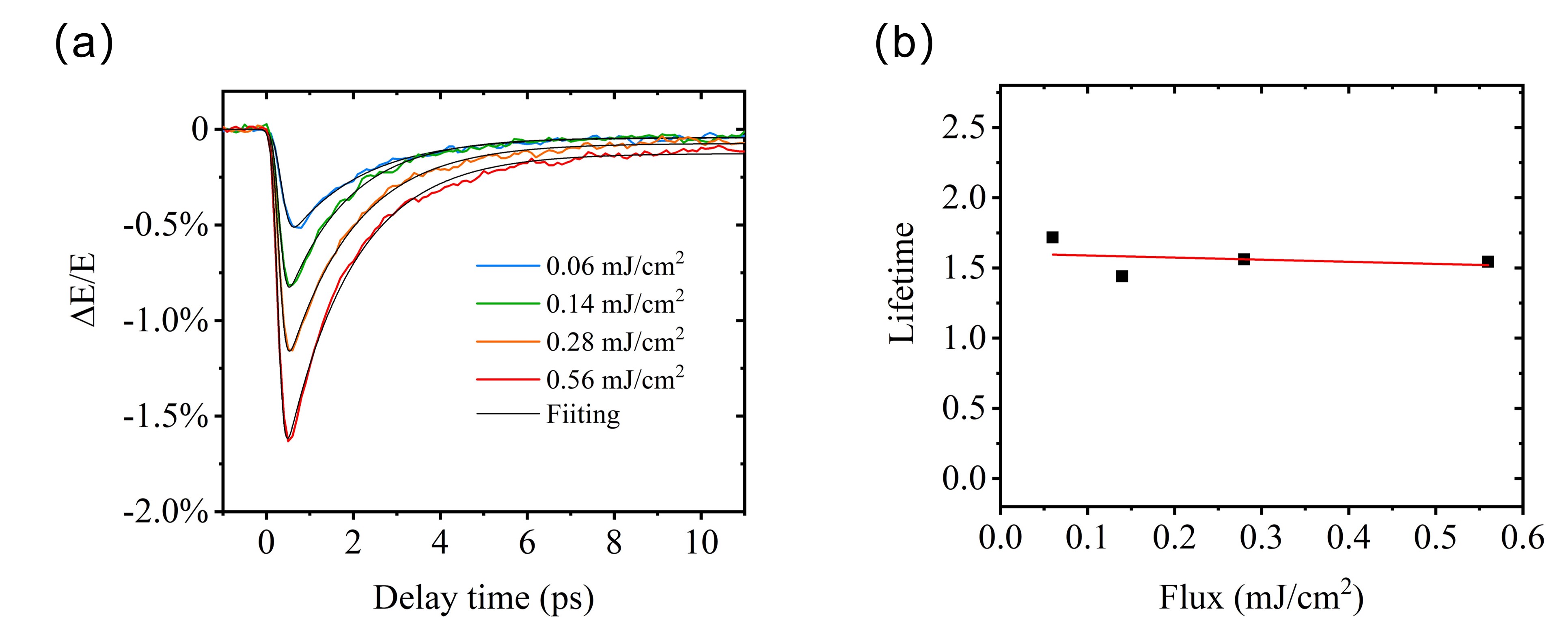}\\
	Fig. S6: (a) The fitting results at different fluences. (b) Lifetime at different fluences.
\end{figure}

\end{appendices}

\begin{center}
\small{\textbf{ACKNOWLEDGMENTS}}
\end{center}

This work was supported by the National Natural Science Foundation of China (Grant No. 11888101), the National Key
Research and Development Program of China (Grant No. 2022YFA1403901).

\bibliographystyle{apsrev4-1}
\bibliography{Ref}

\end{document}